\documentclass[pra,superscriptaddress,12pt,tightenlines]{revtex4}

\usepackage{bm}

\newcommand{\ket}[1]{|#1\rangle}
\newcommand{\proj}[1]{|#1\rangle\langle#1|}
\newcommand{\bra}[1]{{\langle#1|}}

\newcommand{\inner}[2]{{\langle#1|#2\rangle}}
\newcommand{\vc}[2]%
{\left(\begin{array}{c}{\!\!#1\!\!}\\{\!\!#2\!\!}\end{array}\right)}
\newcommand{\D}{\displaystyle}

\def\H{{\cal H}}
\def\HN{\H^{\otimes N}}
\def\Hinf{{\cal H}^{\otimes\infty}}
\def\Hcomp#1{\Hinf_{\{#1\}}}

\def\infvecseq#1{|#1_1\rangle,|#1_2\rangle,\ldots}
\def\infprod#1{|#1_1\rangle\otimes|#1_2\rangle\otimes\cdots}
\def\seq#1{\{#1\}}
\def\seqvec#1{|\seq#1\rangle}
\def\sl#1#2{|#1;\!\seq#2\rangle}
\def\pl#1#2{|#1,#2_1\rangle\otimes|#1,#2_2\rangle\otimes\cdots}
\def\dmu{d\mu_{\sl{\psi}{i}}(\seq j)}
\def\dnu{d\nu_{\,\sl{\psi}{i}}(\seq j)}

\begin{document}

\title{Properties of the frequency operator do not imply the
quantum probability postulate}

\author{Carlton M.~Caves}
\email{caves@info.phys.unm.edu}
\affiliation{Department of Physics
and Astronomy, University of New Mexico, Albuquerque, NM~87131-1156,
USA}
\author{R\"udiger Schack}
\email{r.schack@rhul.ac.uk}
\affiliation{Department of Mathematics, Royal Holloway, University of
London, Egham, Surrey TW20$\;$0EX, UK}

\date{\today}

\begin{abstract}
  We review the properties of the frequency operator for an infinite
  number of systems and disprove claims in the literature that the
  quantum probability postulate can be derived from these properties.
\end{abstract}

\maketitle

\section{Introduction}

Although quantum theory is thought by many physicists to provide a
complete description of the physical world, the account it gives is
strange and counter-intuitive.  Nobody can claim fully to understand
the quantum description, yet we find it appealing because it provides
exquisitely precise predictions for the results of experiments.

The counter-intuitive description provided by quantum mechanics would
be more palatable were it the consequence of a set of compelling,
physically motivated assumptions about the way the world works.  In
contrast to this desire, however, the postulates usually given for
quantum theory are notable for their abstract, mathematical character.
As an illustration, consider how the theory's foundations are
introduced in one standard graduate-level textbook \cite{CT}.  A
chapter of 54 pages is devoted to stating and explaining five
foundational postulates, paraphrased as follows: (i)~the state of a
physical system is a normalized vector $|\psi\rangle$ in a Hilbert
space ${\cal H}$; (ii)~every measurable quantity is described by a
Hermitian operator (observable) $A$ acting in ${\cal H}$; (iii)~the
only possible result of measuring a physical quantity is one of the
eigenvalues of the corresponding observable $A$; (iv)~the probability
for obtaining eigenvalue $\lambda$ in a measurement of $A$ is ${\rm
Pr}(\lambda)=\langle\psi|P_\lambda|\psi\rangle$, where $P_\lambda$ is
the projector onto the eigensubspace of $A$ having eigenvalue
$\lambda$; and (v)~the post-measurement state in such a measurement is
$P_\lambda|\psi\rangle/\sqrt{{\rm Pr}(\lambda)}$.  Compared to the crisp
postulates of special relativity---the laws of physics and the speed of
light are the same in all inertial frames---which are physically
motivated and stated directly in terms of physical concepts and
quantities, the quantum postulates make up a baggy set that can only be
described as more mathematical than physical (see, however,
Refs.~\cite{Hardy2002a,Hardy2002b,Schack2003a} for a derivation of
quantum mechanics from axioms for states and measurements that do not
presuppose a Hilbert-space structure and
Refs.~\cite{Clifton2003,Bub2004a,Bub2004b} for a derivation that places
quantum theory on a foundation of information-theoretic postulates).  A
skeptic, on being exposed to the quantum postulates, would balk after
just the first few of the 54 pages and question the entire mathematical
construction:  Who ordered the complex vector space, which seems to
have nothing to do with the arena of ordinary experience?  How could
anyone think that states of physical systems are vectors in this
abstract space?  How could anyone think that observables are operators
in this space?

One avenue to enlightenment might be to reduce the number of
postulates.  A signal result of this sort, of which we are particularly
fond, is provided by Gleason's theorem~\cite{Gleason1957}.  The theorem
assumes postulates~(ii) and (iii) and that the task of the theory is to
provide probabilities for measurement outcomes, which in accordance
with~(iii), are associated with complete, orthogonal sets of projection
operators.  The key assumption is that these probabilities are {\it
noncontextual\/}~\cite{Caves2002a}, which means that the probability
associated with a projection operator is independent of which other
projectors complete the set of outcomes.  Put differently,
noncontextuality is the assertion that if two observables share an
outcome, i.e., have a shared eigensubspace, then the probability
associated with this outcome is the same for both observables.  The
content of Gleason's theorem is that in Hilbert-space dimensions
$\ge3$, probabilities that are noncontextual must be derived from a
density operator using the mixed-state generalization of the quantum
probability rule in~(iv).  This is particularly pleasing, since it gets
the state-space structure of (i), generalized to density operators, and
the quantum probability rule of (iv), both from the one assumption of
noncontextuality for probabilities of the allowed outcomes in~(iii).
Gleason's theorem doesn't answer the skeptic's fundamental
question---who ordered the complex vector space---but it does suggest
that we can focus our attention on measurements, trying to figure out
why they are described in a Hilbert space, and we can let states come
along for the ride.

Another approach to reducing the number of postulates has been to try
to derive the quantum probability rule~(iv) from the frequency
properties of repeated measurements of an observable on a finite or
infinite number of copies of a system, where all copies are in the same
state.  A critical analysis of programs of this sort is the subject of
this paper.  Such programs are the quantum analogue of the classical
attempt to {\it define\/} probabilities as the frequencies of outcomes
in a finite or infinite number of trials on identical systems. There
are many problems with this frequentist approach to defining
probabilities, and these problems are succinctly summarized in
Ref.~\cite{Appleby2004}.  Ultimately, the program comes down to an
attempt to define probabilities using the weak or strong law of large
numbers.  The program founders because the laws of large numbers
are statements {\it within\/} probability theory, which cannot even be
formulated without reference to probabilities.  For this reason,
attempts to define probabilities using the laws of large numbers are
inherently circular.  The laws of large numbers are indeed important
mathematical results that connect frequencies to probabilities, but
their form illustrates the crucial point: Inferences always run not
from frequencies to probabilities, but from probabilities to
statistical properties of frequencies.

There is, however, some reason to hope that the frequentist program can
be salvaged within a quantum-mechanical framework, because of the
Hilbert-space setting of quantum theory.  The hope is that the
Hilbert-space inner product provides additional structure, not
available in the classical setting, that allows one to make statements
about repeated measurements on an infinite number of copies, statements
that are independent of the quantum probability rule relating inner
products to probabilities.  The mathematical object that embodies this
hope is the frequency operator associated with repeated measurements of
an observable, and the hope has motivated a number of
researchers~\cite{Finkelstein1963,Hartle1968,Farhi1989a,Gutmann1995} to
investigate properties of the frequency operator.

The program followed by these investigators is to consider an infinite
number of copies of a quantum system, all in the same state
$|\psi\rangle$ and all subjected to a measurement of the same
observable.  The chief technical object of the program is to
demonstrate that the resulting infinite repetition state,
$\ket{\Psi_\infty}
=\ket\psi^{\otimes\infty}\equiv\ket\psi\otimes\ket\psi\otimes\cdots\;$,
is an eigenstate of the frequency operator associated with a particular
measurement outcome, with the eigenvalue given by the absolute square
of the inner product for that outcome.  This object accomplished, the
program invokes the rule that when an observable is measured on a
system that is in an eigenstate of the observable, the result is
guaranteed to be the corresponding eigenvalue.  The conclusion is that
an infinite number of measurements of an observable on an
infinite-repetition state yields each outcome with a frequency given by
the corresponding inner product.  By identifying long-run frequencies
with probabilities, this is then interpreted to mean that the
probability for each outcome is given by the appropriate inner product.

The proponents of this program portray it as replacing the quantum
probability rule~(iv) with a weaker hypothesis, which makes no
reference to probabilities, that hypothesis being that a measurement of
an observable on a system in an eigenstate of the observable yields the
eigenvalue with certainty.  It is useful to formalize these two
postulates for later reference in the paper.  The standard quantum
probability postulate has the following form.

\begin{quote}
{\bf Quantum Probability Postulate (QPP)}: Let $B=\sum_\lambda
\lambda P_\lambda$ be an observable, where $\lambda$ denotes the
different eigenvalues of $B$ and the operators $P_\lambda$ are
orthogonal projectors onto the eigenspaces of $B$.  If $B$ is measured
on a system in state $\ket\psi$, the probability of outcome $\lambda$
is $||P_\lambda\ket\psi||^{\,2}=\langle\psi|P_\lambda\ket\psi$.
\end{quote}

\noindent
Here $||\cdot||$ denotes the Euclidean norm, i.e.,
$|||\psi\rangle||=\sqrt{\inner\psi\psi}$.  Throughout this paper, it is
sufficient for us to deal with repeated measurements of a {\it
nondegenerate\/} observable $B$.  In this case, we denote the
eigenvalues of $B$ by $\lambda_j$ and the corresponding eigenvectors by
$|B,j\rangle$, i.e., $B|B,j\rangle=\lambda_j|B,j\rangle$; the
probability for obtaining outcome $\lambda_j$ in a measurement of $B$
becomes $|\langle B,j|\psi\rangle|^2$.  The aim of the frequentist
program is to replace QPP with the weaker postulate of definite
outcomes.

\begin{quote}
{\bf Postulate of Definite Outcomes (PDO)}: If an observable $O$ is
measured on a system in an eigenstate $\ket\psi$ of $O$, i.e.,
$O\ket\psi=\lambda\ket\psi$, the outcome is $\lambda$ with certainty.
\end{quote}

The purpose of this paper is to subject the quantum frequentist program
to critical analysis.  As noted above, the proponents of the program
see their main task as establishing a technical mathematical property
of the frequency operator for an infinite number of copies of a system,
to wit, that an infinite repetition state is an eigenstate of the
infinite-copy frequency operator associated with eigenvalue $\lambda_j$
of observable $B$, with the eigenvalue given by $|\langle B,j|\psi|^2$.
This technical demonstration is indeed at the heart of the program, and
its analysis occupies the main part of our paper; we show that the
desired property of the frequency operator follows only if one assumes
QPP from the outset.  At the end of the paper, we go beyond this
technical result to analyze other aspects of the program, and we find
the program to be flawed at every step.  Our conclusion is that the
inner-product structure of quantum mechanics does not buy additional
power for defining probabilities in terms of infinite frequencies. Even
in quantum mechanics, inferences run from probabilities to frequencies,
not the other way around.

The paper is organized as follows.  Following
Finkelstein~\cite{Finkelstein1963} and Hartle~\cite{Hartle1968},
Sec.~\ref{sec:finiteconstruct} defines the finite-copy frequency
operator $F^N$ and reviews the Finkelstein-Hartle theorem, a
probability-independent statement about the frequency operator that is
mathematically akin to the classical weak law of large numbers.
Section~\ref{sec:finiteStatus} argues that the Finkelstein-Hartle
theorem cannot be used to establish properties of the infinite-copy
frequency operator $F^\infty$, as was claimed by Finkelstein and
Hartle, and thus does not show that infinite-repetition states are
eigenstates of $F^\infty$.

Section~\ref{sec:infinite} takes up the definition of the infinite-copy
frequency operator $F^\infty$.  Following Farhi, Goldstone, and
Gutmann~\cite{Farhi1989a}, Sec.~\ref{sec:Hspace} reviews the
construction of the nonseparable infinite-copy Hilbert space, and then
Sec.~\ref{sec:Gutmann} uses a method due to Gutmann~\cite{Gutmann1995}
to define $F^\infty$.  Gutmann's definition assumes QPP and thus
arrives at an $F^\infty$ that has the property that infinite-repetition
states are eigenstates of $F^\infty$, with the eigenvalues given by the
quantum probabilities.  Gutmann's construction makes clear, however,
that the definition of $F^\infty$ depends on a choice of probability
measure for the space of outcome sequences, and we show that
probability measures other than that dictated by QPP lead to other
eigenvalues for the frequency operator.  This shows that the definition
of $F^\infty$ is not unique and thus that properties of $F^\infty$
cannot be used to pick out the quantum probability rule.  In
Sec.~\ref{sec:Farhi} we examine a derivation of $F^\infty$ due to
Farhi, Goldstone, and Gutmann~\cite{Farhi1989a}, which purports to
derive the unique $F^\infty$ that is consistent with QPP solely from
the inner-product structure.  Cassinello and S\'anchez-G\'omez [15]
criticized this derivation, but their criticism turns out not to be
justified.  We point out in Sec.~IIIC that though the derivation does
pick out a unique measure, other measures can be used to define
$F^\infty$.  Section~\ref{sec:infiniteStatus} shows that an assumption
of noncontextual infinite frequencies is both necessary and sufficient
for picking out the unique $F^\infty$ that is consistent with QPP, thus
making the entire quantum frequentist program merely an elaborate
device for placing the Gleason derivation of QPP within the unnecessary
context of infinite frequencies.

In Sec.~\ref{sec:flaw} we grant the proposition that there is a unique
frequency operator $F^\infty$ such that $\ket\psi^{\otimes\infty}$ is
an eigenstate of $F^\infty$ with eigenvalue $|\langle B,j|\psi\rangle|^2$, but
argue that this cannot be used to conclude anything about single-copy
probabilities.  In Sec.~\ref{sec:certainty}, we point out that the
eigenvalue equation for $F^\infty$ is really a probability-1 statement
and that probability 1 does not imply certainty in uncountable sample
spaces---thus PDO does not hold in nonseparable Hilbert spaces---and
in Sec.~\ref{sec:tailprop} we discuss how tail properties such as the
limiting frequency of an outcome sequence are irrelevant to probabilities
for a finite number of copies.

Section~\ref{sec:conclusion} summarizes our findings and their
implications for the frequentist program to derive the quantum
probability rule.

Throughout the paper we use ``copies'' to refer to multiple versions of
the same quantum system.  Thus we talk about having a finite or
infinite number of copies of a system.  For example, in the former case
we refer to the finite-copy Hilbert space ${\cal H}^{\otimes N}$ and
the finite-copy frequency operator $F^N$ of $N$ copies; likewise, in
the latter case we have the infinite-copy Hilbert space $\Hinf$ and the
infinite-copy frequency operator $F^\infty$.  We reserve the words
``repeated'' and ``repetition'' to refer to situations where the same
state or the same measurement applies to all of the copies, finite or
infinite.  Thus we talk about the repetition state
$\ket{\Psi_N}=\ket\psi^{\otimes N} \equiv
\ket\psi\otimes\cdots\otimes\ket\psi$ of $N$ copies, where
$|\psi\rangle$ is a single-copy state, the infinite-repetition
state $\ket{\Psi_\infty}
=\ket\psi^{\otimes\infty}\equiv\ket\psi\otimes\ket\psi\otimes\cdots\;$
of an infinite number of copies, and repeated measurements of an observable
$B$ on a finite or infinite number of copies.

\section{Frequency operator: finite number of copies}  \label{sec:finite}

\subsection{Construction of the finite-copy frequency operator}
\label{sec:finiteconstruct}

Consider $N$ copies of a quantum system, each copy described in the
same $D$-dimensional Hilbert space ${\cal H}$.  Suppose we measure the
same observable, $B$, on each of the $N$ systems.  We assume for
simplicity that the eigenvalues of $B$ are nondegenerate, although none
of our conclusions depends on this assumption. We denote by $\lambda_j$
the eigenvalues of $B$ and by $\ket{B,j}$ the corresponding
eigenvectors, i.e.,
\begin{equation}
B\ket{B,j}=\lambda_j\ket{B,j}\;,\quad j=0,\ldots,D-1.
\end{equation}
We now single out the outcome $j=0$ (i.e., the eigenvalue $\lambda_0$)
as the outcome of interest.  We are not interested in which of the
other outcomes occurs nor in the order of the outcomes, only in the
number of times, $n$, that outcome $j=0$ occurs.  The frequency of
outcome $j=0$ is the fraction $n/N\equiv f$.

To describe a repeated measurement of $B$ as a single measurement, we
introduce the tensor-product Hilbert space ${\cal H}^{\otimes
N}\equiv{\cal H}\otimes\cdots\otimes{\cal H}$ for $N$ copies of the
system. The projector corresponding to the sequence $j_1,\ldots,j_N$ of
measurement outcomes is given by the tensor product
$\proj{B,j_1}\otimes\cdots\otimes\proj{B,j_N}$. The corresponding
frequency of outcome $j=0$~is
\begin{equation}
f={1\over N}\sum_{r=1}^N\delta_{0j_r}\;.
\end{equation}
Of course, many different outcome sequences give rise to the same
frequency $f$ or occurrence number $n=Nf$.  Adding all the projectors
that lead to this frequency gives the multi-dimensional projector onto
the subspace corresponding to this frequency (or occurrence number):
\begin{eqnarray}
\Pi_n^N&=&
\sum_{j_1,\ldots,j_N}
\proj{B,j_1}\otimes\cdots\otimes\proj{B,j_N}\;
\delta\biggl(n,\sum_{r=1}^N \delta_{0j_r}\biggr)\nonumber\\
&=&\sum_{k_1,\ldots,k_N\in\{0,1\}} P^1_{k_1}
 \otimes\cdots\otimes  P^N_{k_N} \,
 \delta\biggl(n,\sum_{r=1}^N \delta_{0k_r}\biggr) \;.
\label{eqA:Pin}
\end{eqnarray}
Here $P_0^r = \proj{B,0}$, $P_1^r=1-P_0^r$, and the superscripts,
$r=1,\ldots,N$, label which copy the projector applies to.  The
projectors $P_0^r$ and $P_1^r$ make up a binary POVM that describes
retaining only the information about whether the $r$th measurement
yields outcome $j=0$, i.e., $k_r=0$, or some other outcome, i.e.,
$k_r=1$.

The projectors $\Pi_n^N$ clearly add up to the unit operator,
$1^{\otimes N}$, on ${\cal H}^{\otimes N}$ and thus form a POVM made up
of orthogonal multi-dimensional projectors.  This POVM describes a
measurement that can be realized by making $N$ successive measurements
of $B$, after which the outcomes $j\ne0$ are placed in a single bin and
information about the ordering of the outcomes is discarded, leaving
only the outcome frequencies for $j=0$ as measurement results.

By associating the appropriate outcome frequency with each projector
$\Pi_n^N$, the same measurement can be described as a measurement of a
frequency observable, which is the finite-copy {\it frequency
operator\/} for outcome $j=0$,
\begin{equation}
 F^N=
\sum_{n=0}^N{n\over N}\,\Pi^N_n\;.
\label{eqA:freqop}
\end{equation}
Measuring the frequency operator is equivalent to measuring the POVM
consisting of the projection operators~(\ref{eqA:Pin}).

The frequency operator~(\ref{eqA:freqop}) can be put in other useful
forms through the following manipulations:
\begin{eqnarray} \label{eqA:freqopFinkelstein}
F^N
&=&{1\over N}\sum_{n=0}^N\;n
\sum_{k_1,\ldots,k_N\in\{0,1\}}
\delta\biggl(n,\sum_{r=1}^N \delta_{0k_r}\biggr)
P^1_{k_1} \otimes\cdots\otimes  P^N_{k_N}\cr
&=&{1\over N}
\sum_{r=1}^N \;\sum_{k_1,\ldots,k_N\in\{0,1\}}\delta_{0k_r} \,
P^1_{k_1} \otimes\cdots\otimes  P^N_{k_N}\cr
&=& {1\over N}\sum_{r=1}^N
1^{\otimes(r-1)}\otimes
 P^r_0\otimes1^{\otimes(N-r)} \cr
&=& {1\over N} ( P^1_0 + \cdots +  P^N_0) \;.
\end{eqnarray}
The second form on the right is the form in which the frequency
operator was introduced by Hartle \cite{Hartle1968}.  In the final
expression on the right, the frequency operator is written as the
average over the $N$ systems of the projectors onto the chosen outcome.
This final form says that to measure the frequency operator for a
particular outcome, one should measure on each copy the observable that
is the projection operator onto the chosen outcome---the POVM for the
$r$th copy consists of the projection operators $P_1^r$ and
$P_0^r$---and then average the results.  The final form is the form of
the frequency operator that comes from the work of Finkelstein
\cite{Finkelstein1963}, although Finkelstein actually dealt with the
average of a general observable, rather than specifically with the
average of a projector onto a particular outcome.

Regardless of which form one prefers for the frequency operator, one
can see that measuring the frequency operator can be accomplished
through repeated measurements on the separate copies, this despite the
joint operators that appear in Eqs.~(\ref{eqA:Pin}) and
(\ref{eqA:freqopFinkelstein}). Failure to appreciate this point has led
to some confusing discussions in the literature \cite{YAharonov2002}.

An important property of the finite-copy frequency operator is provided by
the {\em Finkelstein-Hartle theorem} \cite{Finkelstein1963,Hartle1968}:
Let $\ket\psi\in{\cal H}$ be a single-copy state, and let
\begin{equation}  \label{eq:productstate}
\ket{\Psi_N}=\ket\psi^{\otimes N} \equiv
\ket\psi\otimes\cdots\otimes\ket\psi
\end{equation}
be the corresponding $N$-copy repetition state; then there is a unique
number $q$ such that
\begin{equation}  \label{eqA:FHtheorem}
\lim_{N\to\infty}|| F^N\ket{\Psi_N}- q\ket{\Psi_N}||
   \equiv\lim_{N\to\infty}\Delta_N=0\;,
\end{equation}
namely
\begin{equation}
q = \bra\psi P_0\ket\psi = |\langle{B,0}\ket\psi|^{\,2} \;.
\label{eqA:q}
\end{equation}

If we start from the final form of $F^N$ in
Eq.~(\ref{eqA:freqopFinkelstein}), the proof of the theorem is
straightforward.  Substituting Eq.~(\ref{eqA:freqopFinkelstein}) into
Eq.~(\ref{eqA:FHtheorem}), we obtain
\begin{equation}
\Delta_N^2 = \Big(q-\bra\psi P_0\ket\psi\Big)^2 +
    {\bra\psi P_0\ket\psi(1 - \bra\psi P_0\ket\psi)\over N}
    \;,
\end{equation}
and the Finkelstein-Hartle theorem follows immediately. We can also
start from the form~(\ref{eqA:freqop}) of $F^N$.  Assuming $q$ is given
by Eq.~(\ref{eqA:q}), we see that $\Delta_N^2$ is the variance of a
random variable $f^N$, which has possible values $n/N$, $n=0,\ldots,N$,
and distribution
\begin{equation}   \label{eq:binomial}
\Pr(f^N=n/N)=||\Pi_n^N\ket{\Psi_N}||^{\,2} = {N\choose n}q^n
(1-q)^{N-n}\;, \quad n=0,\ldots, N\;,
\end{equation}
which is a binomial distribution. This variance is
\begin{eqnarray}
||F^N\ket{\Psi_N}- q\ket{\Psi_N}||^{\,2} &=& E\Big((f^N-q)^2\Big) \nonumber \\
&=& \sum_{n=0}^N \left({n\over N}-q\right)^2 ||P_n^N\ket{\Psi_N}||^{\,2} \nonumber \\
&=&{q(1-q)\over N}\to 0
\quad\mbox{as}\quad N\to \infty \;.
\end{eqnarray}
Here $E$ denotes the expectation value with respect to the
distribution~(\ref{eq:binomial}).
Notice that the Finkelstein-Hartle theorem
is a purely mathematical statement, which uses the results of
probability theory without assuming any {\em probabilistic
interpretation\/} of the above expressions.  The interpretation of the
finite-copy frequency operator is the subject of
Sec.~\ref{sec:finiteStatus} below.

To end the current subsection, we draw attention to another simple
consequence of the binomial distribution~(\ref{eq:binomial}), which in
the context of the frequency operator was first pointed out by Squires
\cite{Squires1990}. Squires's observation was that the repetition state
$|\Psi_N\rangle$ becomes orthogonal to all frequency subspaces as $N$
becomes large,
\begin{equation}  \label{eq:Squires}
\max_{0\le n\le N} || \Pi^N_{n}\ket{\Psi_N} ||^{\,2} \to 0 \;\;\mbox{
  as } \; N\to\infty \;,
\end{equation}
except in the trivial cases $q=0$ and $q=1$. This is, of course,
nothing but the fact that the maximum of the binomial distribution
approaches zero as the number of trials increases.

\subsection{Status of the finite-copy frequency operator}
                  \label{sec:finiteStatus}

The Finkelstein-Hartle theorem suggests that the quantum probability
postulate, QPP, follows directly from the Hilbert-space structure.
Indeed, Finkelstein interpreted Eq.~(\ref{eqA:FHtheorem}) to mean that,
for sufficiently large $N$, the $N$-copy repetition state
$\ket{\Psi_N}$, representing a finite ensemble of systems in the same
quantum state, is ``nearly an eigenstate'' of the average operator
$F^N$, with the deviation from being an eigenstate measured by the
error $\Delta_N$ \cite{Finkelstein1963}.  Vague though this assertion
is, one way to see that it is not warranted is to refer to Squires's
observation that in the limit $N\rightarrow\infty$, the repetition
state $\ket{\Psi_N}$ becomes orthogonal to {\em any\/} eigenstate of
the frequency operator.

Independently of Finkelstein, Hartle \cite{Hartle1968} considered an
infinite number of copies and the infinite-repetition state,
\begin{equation}
\ket{\Psi_\infty}
=\ket\psi^{\otimes\infty}\equiv\ket\psi\otimes\ket\psi\otimes\cdots\;,
\end{equation}
and defined the action of an {\em infinite-copy\/} frequency operator,
$F^\infty$, through the limit
\begin{equation}
F^\infty\ket{\Psi_\infty} = \lim_{N\to\infty}(F^N\ket{\Psi_\infty}) \;.
\label{eq:Hartlelimit}
\end{equation}
He then used Eq.~(\ref{eqA:FHtheorem}) to show that the
infinite-repetition state $\ket{\Psi_\infty}$ is an eigenstate of
$F^\infty$ with eigenvalue $q$, i.e.,
\begin{equation}
F^\infty \ket{\Psi_\infty} = q \ket{\Psi_\infty} \;.
\label{eq:Finfeigenstate}
\end{equation}
We must defer a detailed discussion of the flaw in Hartle's approach
till Sec.~\ref{sec:Gutmann}, but for the present we note that the
finite-copy frequency operator does not have a unique extension to the
infinite-copy Hilbert space, so the operator $F^\infty$
in Eq.~(\ref{eq:Hartlelimit})
is not well defined.  More generally, the $N\to\infty$ limit of
$N$-copy expressions is not sufficient to establish the properties of
infinite-copy expressions.

These points were first stated clearly by Farhi, Goldstone, and Gutmann
(p.~370 of Ref.~\cite{Farhi1989a}).  The Finkelstein-Hartle theorem is
about limits of finite-copy quantities, but the finite-copy repetition
state $\ket\psi^{\otimes N}$ is not an eigenstate of the frequency
operator. The result is that finite-copy considerations do not
determine the infinite-copy frequency operator; one must have a
procedure to extend the definition of the frequency operator to the
nonseparable infinite-copy Hilbert space.  In Sec.~\ref{sec:infinite}
we give a thorough discussion of the infinite-copy Hilbert space and
show in detail why the finite-copy analysis does not determine the
action of the infinite-copy frequency operator on infinite-repetition
states.

This situation is reminiscent of the distinction in probability theory
between the weak and strong laws of large numbers.  The weak law of
large numbers is about the $N\to\infty$ limit of probabilities for $N$
trials, where $N$ is finite, just as the Finkelstein-Hartle theorem is
about the $N\to\infty$ limit of a Hilbert-space norm for $N$ copies. In
contrast, the strong law of large numbers is about probabilities for
infinite sequences of trials, just as the putative eigenvalue
equation~(\ref{eq:Finfeigenstate}) is a direct statement about
infinite-repetition states.  The strong law does not follow from the
weak law, nor does the eigenvalue equation~(\ref{eq:Finfeigenstate})
follow from the Finkelstein-Hartle theorem.  Section~\ref{sec:infinite}
describes a quantum version of the strong law of large numbers.

If the postulate QPP is assumed, as in the standard approach to quantum
theory, only finite-copy considerations are needed to establish a tight
connection between single-trial probabilities and frequencies in
repeated measurements.  It follows directly from QPP that the
probability of measuring a frequency $f=n/N$ is given by the binomial
distribution~(\ref{eq:binomial}), which for large $N$ becomes strongly
peaked near the single-trial probability $q$.  For large $N$, the
measured frequency is close to $q$ with probability close to 1. This is
the familiar connection between measured frequency and probability
expressed by the weak law of large numbers. It is important that in the
formulation of the weak law, frequency and probability are two separate
concepts, with probability being the primary concept: a single-trial
probability distribution is assumed to be
given from the start (in quantum mechanics, provided by QPP for
individual copies); it is then shown that the derived probability
distribution for the measured frequency (a random variable) obeys
certain bounds that are called the weak laws of large numbers.

To summarize, the Finkelstein-Hartle theorem establishes a connection
between the squared inner product and the measured frequency only if
the quantum probability postulate QPP is assumed from the outset. To
overcome this problem, Farhi, Goldstone, and Gutmann \cite{Farhi1989a}
proposed to derive, directly from the properties of Hilbert space, a
unique frequency operator $F^\infty$ defined for an infinite number of
copies of the system, i.e., for an infinite number of measurements.  An
analysis of this derivation and closely related work by
Gutmann~\cite{Gutmann1995} is the topic of the next section.

\section{Frequency operator: infinite number of copies}
\label{sec:infinite}

\subsection{Construction of the infinite-copy Hilbert space}
\label{sec:Hspace}

Consider $N$ copies of a system with a $D$-dimensional Hilbert space
$\H$.  The $N$ copies are described by the $D^N$-dimensional
tensor-product Hilbert space $\HN=\H\otimes\cdots\otimes\H$.  As
before, we denote the orthonormal basis of eigenstates of the measured
observable $B$ in $\H$ by $|B,j\rangle$, where $j=0,\ldots,D-1$.  This
basis gives rise to a product orthonormal basis
$\ket{B,j_1}\otimes\cdots\otimes\ket{B,j_N}$ for $\HN$.

We want to construct the infinite tensor-product space
$\Hinf=\H\otimes\H\otimes\cdots\;$.  This is a {\it nonseparable\/}
Hilbert space, meaning that it has an uncountable orthonormal basis.
Even this means more than one might think initially.  The orthonormal
products of eigenstates of~$B$,
\begin{equation}
\sl{B}{j}\equiv\ket{B,j_1}\otimes\ket{B,j_2}\otimes\cdots\;,
\end{equation}
are in one-to-one correspondence with the real numbers in the interval
$[0,1)$ and thus make up an uncountable set of orthonormal vectors.
Though the finite-copy case might lead one to expect these vectors to
span $\Hinf$, they must be augmented by an uncountable number of other
orthonormal vectors to produce a basis for $\Hinf$.

We follow the discussion contained in Secs.~V and IX of
Ref.~\cite{Farhi1989a}, which constructs $\Hinf$ as a direct sum of an
uncountable number of separable Hilbert spaces called {\em components}.
As noted in Ref.~\cite{Farhi1989a}, this construction is ideally suited
to an analysis of infinite frequencies, because questions about
infinite frequencies can be handled within the separate components.  To
begin, let
\begin{equation}
\seq\psi \equiv \infvecseq\psi
\end{equation}
denote an infinite sequence of normalized vectors in $\H$, and let
\begin{equation}
\seqvec\psi \equiv \infprod\psi
\end{equation}
be the corresponding infinite product vector.  The magnitude of the
inner product of two product vectors,
\begin{equation}
|\langle\seq\phi\seqvec\psi|
=\prod_{r=1}^\infty|\langle\phi_r|\psi_r\rangle|\;,
\end{equation}
lies in the interval $[0,1]$.  Indeed, the inner product goes to zero
unless the product vectors have tail sequences that are essentially
identical.

One defines two sequences to be equivalent, written
$\seq\phi\sim\seq\psi$, if there exists $N\ge1$ such that
\begin{equation}
\prod_{r=N}^\infty|\langle\phi_r|\psi_r\rangle|>0\;,
\label{eq:Ninfty}
\end{equation}
which is equivalent to saying the series
\begin{equation}
\sum_{r=N}^\infty\bigl(-\log|\langle\phi_r|\psi_r\rangle|\bigr)
\end{equation}
converges absolutely.  From the properties of absolutely convergent
series, it is easy to see that two sequences are equivalent if and only
if for any $\epsilon>0$, there exists $N\ge1$ such that
\begin{equation}
\prod_{r=N}^\infty|\langle\phi_r|\psi_r\rangle|>1-\epsilon\;.
\end{equation}
Thus two sequences are equivalent if and only if they have tails that
are essentially identical.  It remains to be shown that $\sim$ defines
an equivalence relation, but this is not hard to do, the only part that
requires a little work being transitivity.  Notice that vectors
corresponding to inequivalent sequences are orthogonal.

The {\it component\/} associated with an equivalence class is defined
to be the subspace spanned by the infinite product vectors in the
class. To show that each component is a separable Hilbert space, one
constructs a countable orthonormal basis for the component in the
following way. Select a sequence $\seq\psi=\infvecseq\psi$ from the
equivalence class that defines the component, and call the component
$\Hcomp\psi$. For each vector $|\psi_r\rangle$ in the representative
sequence, choose an orthonormal basis
$\ket{\psi_r,0},\ldots,\ket{\psi_r,D-1}$ such that
$\ket{\psi_r,0}=\ket{\psi_r}$. Now define the sequences $\seq
i=i_1,i_2,\ldots\,$, where $i_k=0,\ldots D-1$, to be those with a {\it
finite\/} number of nonzero elements. These sequences are countable.
The corresponding sequences of vectors,
$\ket{\psi_1,i_1},\ket{\psi_2,i_2}\ldots$, are clearly in the
equivalence class of $\seq\psi$.   What we want to show is that the
corresponding product vectors,
\begin{equation}
\sl{\psi}{i} \equiv \ket{\psi_1,i_1}\otimes\ket{\psi_2,i_2}\otimes\cdots\;,
\end{equation}
span $\Hcomp\psi$.  To do so, one needs to show that the vector
corresponding to any sequence $\seq\phi\sim\seq\psi$ can be expanded in
terms of the vectors $\sl{\psi}{i}$.  The expansion should look like
\begin{equation}
\seqvec\phi=\sum_{\seq i}\ket{\psi;\{i\}}\bra{\psi;\{i\}}\{\phi\}\rangle\;.
\end{equation}
The vectors $\sl{\psi}{i}$ are complete in $\Hcomp\psi$ if and only if
the inner-product expansion coefficients satisfy the completeness
condition,
\begin{equation}
\sum_{\seq i}|\langle\psi;\seq i\seqvec\phi|^{\,2}=1\quad \mbox{for all
$\seq\phi\sim\seq\psi$.}
\label{eq:completeness}
\end{equation}
The inner-product expansion coefficients have the explicit form
\begin{equation}   \label{eq:innerexpansion}
\langle\psi;\seq i\seqvec\phi
    =\prod_{r=1}^\infty\langle\psi_r,i_r|\phi_r\rangle\;.
\end{equation}
Once past the finite number of nonzero entries in $\seq i$, the terms in
Eq.~(\ref{eq:innerexpansion}) are identical to those in
$\langle\seq\psi\seqvec\phi$. By rephasing the vectors $|\phi_r\rangle$
(at the expense of introducing an overall phase into $\seqvec\phi$),
one can make all the inner products $\langle\psi_r|\phi_r\rangle$ real
and nonnegative.  Then the terms in the tail of the inner product
$\langle\psi;\seq i\seqvec\phi$ are just
$|\langle\psi_r|\phi_r\rangle|$.

To demonstrate the completeness condition~(\ref{eq:completeness}), one
begins by noting that for any $\epsilon>0$, there exists $N$ such that
Eq.~(\ref{eq:Ninfty}) is satisfied, which implies that
\begin{equation}
\prod_{r=N+1}^\infty
|\langle\psi_r|\phi_r\rangle|^{\,2}>(1-\epsilon)^2>1-2\epsilon\;.
\end{equation}
Now consider the sequences $\seq i$ such that $i_r=0$ for $r>N$.  These
sequences run over all possibilities in the first $N$ slots and are
completed by 0's in the tail slots $r>N$.  For these sequences, we can
write
\begin{equation}
\langle\psi;\seq i\seqvec\phi=
\prod_{r=1}^N\langle\psi_r,i_r|\phi_r\rangle
\prod_{r=N+1}^\infty|\langle \psi_r|\phi_r\rangle|\;.
\end{equation}
Now summing only over the sequences such that $i_r=0$ for $r>N$ (sum
denoted by a prime), we have
\begin{equation}
{\sum_{\seq i}}'|\langle\psi;\seq i\seqvec\phi|^{\,2}=
\sum_{i_1=0}^{D-1}|\langle\psi_1,i_1|\phi_1\rangle|^{\,2}
\cdots\sum_{i_N=0}^{D-1}\langle\psi_N,i_N|\phi_N\rangle|^{\,2}
\prod_{r=N+1}^\infty|\langle \psi_r|\phi_r\rangle|^{\,2}
>1-2\epsilon\;,
\end{equation}
since the $N$ sums are all equal to 1.  This result implies that the
unrestricted sum converges to 1, completing the demonstration that the
sequences $\sl{\psi}{i}$ span $\Hcomp\psi$.

To summarize, there are an uncountable number of components, each
corresponding to an equivalence class of product states that are
essentially identical in the tail.  Each component is a separable
Hilbert space, spanned by a countable orthonormal basis.  Different
components are orthogonal, and the entire Hilbert space $\Hinf$ is the
direct sum of the components.   We draw attention to the fact that {\it
every\/} component contains the entire Hilbert space $\HN$ for the
first $N$ copies for any finite value of $N$; i.e., in mathematical
language, $\Hcomp{\psi}=\HN\otimes\Hcomp{\psi'}$, where $\seq{\psi'}$
denotes the defining sequence $\seq{\psi}$ with the first $N$ vectors
omitted.  This means that the difference between components lies
entirely in the tails of the equivalence classes defining the
components. The subspace spanned by the product vectors $\sl{B}{j}$ is
already a direct sum of an uncountable number of components, yet there
are an uncountable number of other components.

\subsection{Infinite-copy frequency operator}
\label{sec:Gutmann}

The task now is to define a frequency operator $F^\infty$ on the
infinite-copy Hilbert space $\Hinf$.  In doing so, the finite-copy
frequency operator $F^N$ is of little help.  All that $F^N$ tells us is
how to define the action of $F^\infty$ on the products of eigenstates,
$\sl{B}{j}=\ket{B,j_1}\otimes\ket{B,j_2}\otimes\cdots\,$, but this
doesn't go very far, because there is an uncountable number of other
components of $\Hinf$ on which we still have to define $F^\infty$.  To
define $F^\infty$, we must extend its action to these other components,
containing states like the infinite-repetition states
$|\Psi_\infty\rangle$, which are our main interest.  The treatment of
the extension in this subsection follows closely the account given by
Gutmann~\cite{Gutmann1995}.

We begin by defining the frequency of a sequence, $\seq j$, to be
\begin{equation} \label{eq:freqdef}
f(\seq j)={1\over2}
\left(\limsup_{N\rightarrow\infty}{1\over N}\sum_{r=1}^N \delta_{0j_r}
+\liminf_{N\rightarrow\infty}{1\over N}\sum_{r=1}^N \delta_{0j_r}\right)\;,
\end{equation}
where, for a sequence $a_N$,
$\D{\liminf a_N =
  \lim_{N\to\infty}(\inf_{k\ge N} a_k)}$ and $\D{\limsup a_N =
  \lim_{N\to\infty}(\sup_{k\ge N} a_k)}$. If there is a limiting
frequency, we can dispense with the $\limsup$'s and $\liminf$'s,
writing
\begin{equation}
f(\seq j)=\lim_{N\rightarrow\infty}{1\over N}\sum_{r=1}^N \delta_{0j_r}\;.
\end{equation}

We need the following generalization of the strong law of large
numbers, which is a simple consequence of Theorem~3 in Sec.~VII.8 of
Ref.~\cite{Feller1971}.  Let $\seq q=q_1,q_2,\ldots$ be an arbitrary
sequence of probabilities, i.e., $0\le q_r\le1$ for all $r$, and let
$X_1,X_2,\ldots$ be a sequence of independent binary random variables
such that $X_r\in\{0,1\}$ and ${\rm Pr}(X_r=0)=q_r$ for all $r$. Define
the average probability
\begin{equation}    \label{eq:fellerTheorem}
f_{\seq q} = {1\over2}
\left(\limsup_{N\rightarrow\infty}{1\over N}\sum_{r=1}^N q_r
+\liminf_{N\rightarrow\infty}{1\over N}\sum_{r=1}^N q_r \right)
\end{equation}
and the frequency random variable
\begin{equation}    \label{eq:fellerTheorem2}
f^\infty = {1\over2}
\left(\limsup_{N\rightarrow\infty}{1\over N}\sum_{r=1}^N \delta_{0X_r}
+\liminf_{N\rightarrow\infty}{1\over N}\sum_{r=1}^N \delta_{0X_r}\right)\;,
\end{equation}
where $\delta_{0X_r}=1-X_r$.  Then $f^\infty=f_{\seq q}$ with
probability 1.

Consider now a component represented by a sequence $\seq\psi$.  As
discussed in the preceding subsection, the component $\Hcomp\psi$ is
spanned by a countable orthonormal basis $\sl{\psi}{i}=\pl{\psi}{i}\,$,
where the sequences $\seq i$ have only a finite number of nonzero
entries. Following Gutmann~\cite{Gutmann1995}, we now associate  with a
given state $\sl{\psi}{i}$ a probability measure, $\dnu$, on the space
of infinite sequences $\seq j$ of outcomes for repeated measurements of
$B$.  Gutmann chooses the measure associated with all sequences
beginning with $j_1,\ldots,j_N$ to be that given by the quantum
probability rule,
\begin{equation}     \label{eq:Gthree}
\nu_{\,\sl{\psi}{i}}(j_1,\ldots,j_N) =
\int d\nu_{\,\sl{\psi}{i}}(\seq{j^{\,\prime}})\,\prod_{r=1}^N \delta_{j_rj_r'}
=\prod_{r=1}^N q_{\ket{\psi_r,i_r}}(j_r) \;,
\end{equation}
where
\begin{equation} \label{eq:qr}
q_{\ket{\psi_r,i_r}}(j_r)=|\langle\psi_r,i_r|B,j_r\rangle|^{\,2}\;.
\end{equation}
These $N$-copy conditions (for all $N$) determine the measure
$d\nu_{\,\sl{\psi}{i}}$.

Two points should be emphasized here.  First, if $\seq\psi$ is not
equivalent to any of the products of eigenstates of $B$ (i.e., the
states $\sl{B}{j}$), then the product of inner products in
Eq.~(\ref{eq:Gthree}) goes to zero as $N$ goes to infinity, because
vectors from different components are orthogonal.  This is essentially
Squires's observation, and it means only that the probability for any
infinite sequence is zero.  Second, the measure $\dnu$ depends on the
particular state $\sl{\psi}{i}$. This must be the case in order to get
the ``right measure'' for the initial part of a sequence $\seq j$.
Nevertheless, since all the sequences $\seq i$ have the same tails,
consisting entirely of zeroes, the tail terms in the above product of
inner products are independent of $\seq i$, given by
$|\langle\psi_r|B,j_r\rangle|^{\,2}$.  A function of outcome sequences
whose value is determined by the tail of $\seq j$, i.e., is independent
of $j_1,\ldots,j_N$ for any finite value of $N$, is called a {\it tail
property\/}~\cite{Feller1971}.  When integrating a tail property over
$\dnu$, all the measures give the same result, independent of $\seq i$.

An example of a tail property is the frequency~(\ref{eq:freqdef}) of
outcome~$j=0$.  The average frequency is given by the integral
\begin{equation}
\int\dnu\,f(\seq j)=f_{\seq q} \;,
\end{equation}
where $f_{\seq q}$ is the average probability of
Eq.~(\ref{eq:fellerTheorem}), with the sequence of probabilities, $\seq
q$, given by Eq.~(\ref{eq:qr}) with $j_r=0$, i.e.,
\begin{equation}
q_r=\int \dnu\delta_{0j_r}
=|\langle\psi_r,i_r|B,0\rangle|^{\,2}
=q_{\ket{\psi_r,i_r}}(0)\;.
\label{eq:qr2}
\end{equation}
Once past the finite number of nonzero entries in $\seq i$, the tail
terms in $\seq q$ are independent of $\seq i$, given by
$q_r=|\langle\psi_r|B,0\rangle|^{\,2}$.  The frequency of a sequence is
determined by the tail, so the limit for $f_{\seq q}$ is independent of
$\seq i$.  Thus we write
\begin{equation}
f_{\seq q} = f_{\seq\psi} \;,
\end{equation}
emphasizing that this is a unique frequency associated with the
component $\Hcomp\psi$.

We now define the projector $\Pi_f^\infty$ that projects onto frequency
$f$.  Following Gutmann, we define the action of $\Pi_f^\infty$ on the
component $\Hcomp\psi$ by requiring
\begin{equation}   \label{eq:Gone}
\bigl|\bigl|\Pi_f^\infty\sl{\psi}{i}\bigr|\bigr|^{\,2}
=\int\dnu\,\Pi_f(\seq j)\;,
\end{equation}
where
\begin{equation}    \label{eq:Gtwo}
\Pi_f(\seq j)=\cases{1\;,&if $f(\seq j)=f$,\cr
                     0\;,&if $f(\seq j)\ne f$,}
\end{equation}
is the indicator function for frequency $f$, meant to characterize the
desired properties of the projection operator.  The generalization of
the strong law of large numbers quoted above gives us immediately that
\begin{equation}
\bigl|\bigl|\Pi_f^\infty\sl{\psi}{i}\bigr|\bigr|^{\,2}
=\int\dnu\,\Pi_f(\seq j) =\cases{1\;,&if $f=f_{\seq\psi}$,\cr
        0\;,&if $f\ne f_{\seq\psi}$.}
\label{eq:quantumslln}
\end{equation}
We can thus proceed to define
\begin{equation}
\Pi_f^\infty\sl{\psi}{i}=
\cases{\sl{\psi}{i}\;,&if $f=f_{\seq\psi}$,\cr
       0\;,&if $f\ne f_{\seq\psi}$.}
\end{equation}
Since the vectors $\sl{\psi}{i}$ span $\Hcomp\psi$, this can be
extended to all vectors $|\Psi\rangle\in\Hcomp\psi$,
\begin{equation}
\Pi_f^\infty|\Psi\rangle=
\cases{|\Psi\rangle\;,&if $f=f_{\seq\psi}$,\cr
       0\;,&if $f\ne f_{\seq\psi}$.}
\end{equation}
This result is a quantum version of the strong law of large numbers,
following directly from the classical strong law in the form expressed
above.  It is clear now that we can define the infinite-copy frequency
operator by
\begin{equation}
F^\infty|\Psi\rangle=f_{\seq\psi}|\Psi\rangle \;.
\end{equation}
Each component is an eigensubspace of the infinite-copy frequency
operator; i.e., all vectors in a component are eigenvectors of
$F^\infty$, all having the same frequency eigenvalue.

All of this simplifies in the situation of most interest, where the
component under consideration has a representative sequence that
consists of identical vectors, i.e.,
$\seq\psi=|\psi\rangle,|\psi\rangle,\ldots\;$. Then the tail of $\seq
q$ is independent of $r$, i.e., $q_r=|\langle\psi|B,0\rangle|^{\,2}$,
and the frequency associated with $\Hcomp\psi$ is
$f_{\seq\psi}=|\langle\psi|B,0\rangle|^{\,2}$.

It is instructive to pause here and ponder what all this means.
Equation~(\ref{eq:Gone}) equates two ways of writing the probability
for frequency $f$ in an infinite number of measurements of $B$ on the
product state $\sl{\psi}{i}$.  The indicator function~(\ref{eq:Gtwo})
restricts the integral on the right of Eq.~({\ref{eq:Gone}) to
sequences of measurement outcomes that have a particular frequency $f$,
so the integral reports the probability for finding that frequency,
just what the matrix element on the left of Eq.~(\ref{eq:Gone}) is
supposed to be.  The strong law of large numbers is invoked to evaluate
the integral as being either 0 or 1, thus defining the frequency
projection operators.  The definition is determined by the measure
$\dnu$ of Eq.~(\ref{eq:Gthree}), which is the unique choice if one has
already identified the absolute square of inner products with
probabilities.  This is all Gutmann~\cite{Gutmann1995} has in mind,
since he is interested not in deriving the quantum probability rule,
but rather in defining projection operators on $\Hinf$ and deriving
properties of these operators using the strong law of large numbers for
the quantum probabilities.

On the other hand, if one is trying to {\it derive\/} the quantum
probability rule from the result that $\sl{\psi}{i}$ is an eigenstate
of $F^\infty$ with the ``right'' eigenvalue, then one needs to think
harder about the procedure used to get to this result.  The starting
point is Eq.~(\ref{eq:Gone}), which says that the overlap of
$\sl{\psi}{i}$ with the subspace of sequences with frequency $f$ is to
be identified with the integral of the indicator function $\Pi_f(\seq
j)$ over the quantum probability measure $\dnu$.  The measure is
clearly the whole story, and one has to justify the choice of the
quantum probability measure in the absence of any {\it a priori\/}
connection between inner products and probabilities.

One might argue that $\dnu$ is the only possible measure, given the
inner-product structure on Hilbert space, but it is easy to see that
there are other choices.  Suppose, for example, that one adopts a
measure $\dmu$ specified by
\begin{equation}     \label{eq:newmu1}
\mu_{\sl{\psi}{i}}(j_1,\ldots,j_N) =\int
d\mu_{\sl{\psi}{i}}(\seq{j^{\,\prime}})\,\prod_{r=1}^N \delta_{j_rj_r'}
=\prod_{r=1}^N q_{\ket{\psi_r,i_r}}(j_r) \;,
\end{equation}
where the terms in the product are now given by
\begin{equation}      \label{eq:newmu2}
q_{\ket{\psi_r,i_r}}(j_r)=
{\cal N}_r^{-1}\,g(|\langle\psi_r,i_r|B,j_r\rangle|)\;.
\end{equation}
Here $g$ is a function that maps the interval $[0,1]$ to itself,
satisfying $g(0)=0$ and $g(1)=1$, and the normalization factor ${\cal
N}_r$ is given by
\begin{equation}
{\cal N}_r= \sum_{j_r=0}^{D-1}g(|\langle\psi_r,i_r|B,j_r\rangle|) \;.
\end{equation}
The standard quantum measure corresponds to $g(x)=x^2$, but we can
equally well use any other function, such as $g(x)=x^4$, if we don't
already have the quantum probability rule in mind.  With the measure
specified by Eq.~(\ref{eq:newmu1}), Eq.~(\ref{eq:qr2}) is replaced by
\begin{equation}  \label{eq:qrrevised}
q_r=\int\dmu\,\delta_{0j_r}=
{\cal N}_r^{-1}g(|\langle\psi_r,i_r|B,0\rangle|)=
q_{\ket{\psi_r,i_r}}(0)
\;,
\end{equation}
which determines the frequency $f_{\seq\psi}$ associated with
$\Hcomp\psi$.  In the case of a sequence of identical vectors, the
frequency becomes
\begin{equation}
f_{\seq\psi}=
{g(|\langle\psi|B,0\rangle|)\over
\displaystyle{\sum_j g(|\langle\psi|B,j\rangle|)}}\;.
\end{equation}

The upshot of this discussion is that there is no unique extension of
the finite-copy frequency operator to the infinite-copy Hilbert space.
Indeed, there are even more general choices of measure in which one
allows $q_r$ to depend on the phase of the inner product
$\langle\psi_r,i_r|B,0\rangle$, but there is no need for us to consider
these more general measures here.

Another way of illustrating the lack of uniqueness is to consider
infinite tensor products of eigenstates of the measured observable,
i.e., the states $\sl{B}{j}$.  These states being eigenstates of $F^N$,
the finite-copy frequency operator tells us---this is the only thing it
tells us the infinite limit---that these states should be
eigenstates of $F^\infty$, with the eigenvalue given by the
frequency~(\ref{eq:freqdef}) of the binary sequence
$\delta_{0j_1},\delta_{0j_2},\ldots\;$.  To see how this frequency
arises from the measures identified in Eqs.~(\ref{eq:newmu1}) and
(\ref{eq:newmu2}), suppose the component's representative sequence
corresponds to the product state $\seqvec\psi=\sl{B}{j}$.  Then, as a
consequence of the requirements $g(0)=0$ and $g(1)=1$ and nothing else,
no matter what product vector $\sl{\psi}{i}$ is chosen to define the
measure, the sequence of probabilities, $\seq q$, of
Eq.~(\ref{eq:qrrevised}) has a tail that is precisely the required
binary sequence, i.e., $q_r=\delta_{0j_r}$ in the tail.  The
component's frequency $f_{\seq\psi}=f_{\seq q}$ is thus the frequency
of this binary sequence, as required.  To summarize, the only thing
that the finite-copy frequency operator tells us in the infinite limit
is the endpoints of the function $g$, in which case we are dealing
entirely with measurement results that are certainties. The
nonuniqueness of the measure comes from the fact that the behavior of
$g$ away from the endpoints is determined by what we {\em assume\/} for
the single-copy measurement probabilities.

The lack of a unique extension is what dooms Hartle's approach
\cite{Hartle1968}.  In a Mathematical Appendix to his classic 1968
paper, Hartle defines a particular extension $F^\infty$ and proceeds to
show that the infinite repetition states $\ket{\Psi_\infty}$ are
eigenstates of $F^\infty$ with the ``right'' eigenvalue.  The work in
this subsection shows, however, that the extension cannot be unique, so
Hartle's extension contains implicitly an assumption of the quantum
probability rule for infinite repetition states.
We note in addition that the particular operator $F^\infty$
that Hartle defines is not a reasonable extension of $F^N$, because it
is defined to give the ``right'' frequencies only for states in the
symmetric subspace, i.e., states in the subspace spanned by the
infinite-repetition states.  This means, in particular, that Hartle's
extension does not give the right frequencies for infinite tensor
products of eigenstates of the measured observable, frequencies that
are determined uniquely in the limit.

\subsection{Derivation of the infinite-copy frequency operator in
Farhi {\em et al.}}
\label{sec:Farhi}

The importance of Ref.~[13] is that the authors claim to {\it derive\/}
the quantum form of the measure, given in Eqs.~(34) and (35), without
assuming the quantum probability rule, QPP.  This derivation thus
deserves close attention, because it purports to determine the quantum
form of the measure solely from the inner-product structure of $\Hinf$,
as expressed in the properties of unitary transformations between
bases.  This subsection is restricted to the case of a two-dimensional
system, $D=2$, in order to match the analysis in Sec.~V of Ref.~[13].

In Sec.~V of Ref.~[13], the authors set out to construct simultaneous
eigenstates $\sl{\,b}{j}$ of the measured observables,
$B^1,B^2,\ldots\,$, within a component $\Hcomp\psi$ whose
representative sequence, $\seq\psi=|\psi\rangle,|\psi\rangle,\ldots\,$,
consists of identical vectors.  (It is easy to generalize to other
components, but there is no need to do so.)  Here as previously, the
superscript on the measured observable $B$ specifies which copy the
operator acts on.

The states $\sl{\,b}{j}$ are in one-to-one correspondence with the
eigenstate products $\sl{B}{j}$, but they are not the eigenstate
products except in the uninteresting case where $|\psi\rangle$ is one
of the eigenstates of the measured observable. Indeed, the states
$\sl{\,b}{j}$ are unnormalizable, as is pointed out in Ref.~[13] and as
must be true since they are an uncountable basis for the separable
component $\Hcomp\psi$.  The states $\sl{\,b}{j}$ are determined by
defining the inner products $\langle\psi;\seq i\sl{\,b}{j}$, which
allows one to write the states $\sl{\,b}{j}$ in terms of the basis
states $|\psi;\seq{i}\rangle$ for the component in question.  Not
surprisingly, this transformation hinges on a measure $d\mu(\seq j)$
for integrating over the uncountable infinity of unnormalizable states
$\sl{\,b}{j}$.  As shown by Farhi, Goldstone, and Gutmann [13], this
measure is uniquely determined by the inner-product structure of the
component, and hence it is the measure $d\nu_{\seq\psi}(\seq j)$ that
is associated with the standard inner-product quantum probability rule
for the representative sequence $\seq i=\seq 0$.

It is tempting to view the states $\sl{\,b}{j}$ as playing the role of
the eigenstate products $\sl{B}{j}$ and thus as dictating the
definition of $F^\infty$ in the component $\Hcomp\psi$, and this is
what is done in Sec.~VII of Ref.~[13].  This procedure is not
justified, however, because the states $\sl{\,b}{j}$ do not lie in the
component $\Hcomp\psi$---indeed, as unnormalizable states, they do not
lie in the nonseparable infinite-copy Hilbert space $\Hinf$---although
as bras they are legitimate dual vectors for the component.  The
procedure used in Sec.~VII of Ref.~[13] is formally attractive, but it
has no physically motivated justification.  In particular, that the
standard measure emerges from the transformation to the states
$\sl{\,b}{j}$ does not mean that it is the only measure that can be
used to define the infinite-copy frequency operator.  This is clear
from the other measures exhibited in Sec.~IIIB, all of which can be
used to define $F^\infty$.

To think that the measure involved in the transformation between the
states $|\psi;\seq{i}\rangle$ and the unnormalizable states
$\sl{\,b}{j}$ dictates QPP is equivalent to regarding the unitary
transformation between two orthonormal bases as determining QPP.
Unitary transformations between bases are an expression of the
inner-product structure of Hilbert space, since they are the unique
transformations that preserve inner products.  This means that
probabilities derived from QPP transform in a particularly simple way,
when compared to other possible rules, but this is not sufficient to
reject other probability rules without further assumptions.

\subsection{Status of the infinite-copy frequency operator}
\label{sec:infiniteStatus}

The work in this section is devoted entirely to the definition and
mathematical properties of the infinite-copy frequency operator
$F^\infty$.  This work shows convincingly, we think, that absent the
quantum probability rule, there is no justification for choosing the
measure that makes infinite repetition states eigenstates of $F^\infty$
with the frequency eigenvalue given by the quantum probability rule. Indeed,
you can get any frequency eigenvalue you want, unless you have already
assumed the quantum probability rule to fix the choice of measure.

To derive particular eigenvalues for the frequency operator requires
additional assumptions.  Specifically, if one wishes the limiting
frequencies to be those given by the quantum probability rule, then one
of these assumptions must be that of {\em noncontextual limiting
frequencies}, i.e., that the limiting frequency of the selected outcome
(chosen to be $j=0$ in our previous discussion) in repeated
measurements does not depend on the other possible outcomes of the
measurement.  That one must assume noncontextuality or its equivalent
is illustrated by the alternative measures identified in
Eqs.~(\ref{eq:newmu1}) and (\ref{eq:newmu2}).  For all these measures,
except the quantum measure $g(x)=x^2$, the normalization factor ${\cal
N}_r$ in the quantities $q_r$ of Eq.~(\ref{eq:qrrevised}) means that
for dimensions $D\ge3$, these quantities and, hence, the limiting
frequencies are contextual, depending on eigenstates of the measured
observable other than the eigenstate for the selected outcome.

This argument can be put in the formal context of Gleason's theorem
\cite{Gleason1957}.  Suppose one has an infinite number of independent
copies of a quantum system on which one makes repeated measurements of
the observable~$B$.  Without saying anything about the state of the
copies, one can say that the measure and, hence, the limiting frequency
for a selected outcome~$j^{\,\prime}$ is determined by a sequence $\seq
q$, where the entry for the $r$th copy has the form
\begin{equation} \label{eq:framezero}
q_r=q_r(j^{\,\prime};\{\ket{B,j}\})\;,
\end{equation}
signifying that it generally depends both on the set of eigenstates of
the measured observable and on the selected outcome~$j^{\,\prime}$.
Since some outcome occurs for each copy, we must have
\begin{equation} \label{eq:frameone}
\sum_{j^{\,\prime}=0}^{D-1}q_r(j^{\,\prime};\{\ket{B,j}\})=1\;.
\end{equation}
The assumption of noncontextual limiting frequencies asserts that for
repeated measurements of any other observable $C$ that shares an
eigenstate with $B$, i.e., $\ket{C,k'}=\ket{B,j^{\,\prime}}$ for some $k'$, we must
have
\begin{equation} \label{eq:frametwo}
q_r(j^{\,\prime};\{\ket{B,j}\})=q_r(k^{\,\prime};\{\ket{C,k}\})\;.
\end{equation}
This means that $q_r$ depends only on the eigenstate of the selected
outcome and not on the other eigenstates of the measured observable.
Technically, since limiting frequencies are determined by the tail of
the sequence $\seq q$, Eq.~(\ref{eq:frametwo}) only needs to hold in
the tail, i.e., for copies beyond any finite number.  There being no
difference between the tail copies and the leading copies, however, we
extend Eq.~(\ref{eq:frametwo}) to all copies.  In the jargon of
Gleason's theorem, Eqs.~(\ref{eq:frameone}) and (\ref{eq:frametwo})
mean that $q_r$ is a frame function, i.e., a function on pure states
that sums to a constant (here equal to 1) on orthonormal bases.  Gleason's
theorem~\cite{Gleason1957} implies that in dimensions $D\ge3$, any
such function has the form $q_r=\langle B,j|\rho_r|B,j\rangle$ for
some normalized density operator $\rho_r$.  The infinite product state
$\rho_1\otimes\rho_2\otimes\cdots\,$ thus becomes the state of the
infinite-copy system, and the elements of the sequence $\seq q$ and,
hence, the eigenvalues of $F^\infty$ are computed using the quantum
probability rule.

Although this argument does show that the assumption of noncontextual
limiting frequencies picks out the quantum probability rule, what it
really shows is the bankruptcy of the program of deriving quantum
probabilities from limiting frequencies.  The first step in the
argument, that the limiting frequency is derived from a sequence $\seq
q$, is only justified if one interprets the $r$th element of the
sequence, the quantity $q_r$ of Eq.~(\ref{eq:framezero}), as the
probability for obtaining outcome $j$ in a measurement of $B$ on the
$r$th copy.  With this realization, the elaborate superstructure of
repeated measurements on an infinite number of copies collapses,
revealed as irrelevant to an argument that really deals directly with
single-copy probabilities.  With the superstructure swept away, the
argument stands forth in its original form as the pristine Gleason
derivation of the state-space structure of quantum mechanics and the
quantum probability rule from the assumption of noncontextual
probabilities for quantum measurements \cite{Caves2002a}.

In the end, the point is that infinite frequencies don't determine
probabilities, either classically or quantum mechanically.  Inferences
always flow in the opposite direction: probabilities are the primary
concept, and they determine properties of infinite frequencies.  As
discussed in Sec.~\ref{sec:Gutmann} and illustrated further in this
subsection, the eigenvalues of the infinite-copy frequency operator
depend on what one assumes for single-copy probabilities.  The
inner-product structure of Hilbert space is insufficient by itself to
determine the frequency eigenvalues.  Instead of PDO determining the
single-copy probability rule from the eigenvalues of $F^\infty$, what
is true is that the only eigenvalues of $F^\infty$ that are determined
without reference to single-copy probabilities are those that are
dictated by repeated measurements with definite outcomes.

\section{Deriving QPP from PDO is flawed from the outset}
\label{sec:flaw}

We could stop at the end of the preceding section, having shown that
the eigenvalues of $F^\infty$ are not uniquely determined without
reference to single-copy probabilities, but a critical analysis of the
program to derive QPP from the properties of $F^\infty$ would not be
complete without a discussion of the points in this section.  In this
section, we grant the proposition that there is a unique infinite-copy
frequency operator whose eigenvalues are given by the quantum
probability rule, but we argue that one still cannot derive QPP by
applying PDO to this result.  The two arguments made in this section
are really arguments within classical probability theory.  They can be
made against the classical frequentist program to derive probabilities
from the strong law of large numbers, but here we put these arguments
specifically within the quantum context.

\subsection{Certainty versus probability 1}
\label{sec:certainty}

For finite or countably infinite sample spaces, if a subset has
probability 1, then in selecting an alternative from the sample space,
one of the alternatives in the subset is certain to occur; likewise,
probability 0 means impossibility.  These statements are no longer true
for uncountable sample spaces, where probability 1 does not imply
certainty.  Any alternative or any subset of measure zero in an uncountable
sample space has zero
probability.  Any alternative or any subset of measure zero can be moved
from a set of probability 1 to the complementary subset of probability
0 without changing these probabilities.  If one believed that
probability 0 implied impossibility, one would conclude that all
alternatives were impossible.  An example of this is provided by the
uncountable sample space of outcome sequences for an infinite sequence
of trials, where the strong law of large numbers establishes that with
probability 1 the frequency of occurrence of an outcome is equal to its
probability, but does not imply that the frequency is certain to be
equal to the probability.  The strong law of large numbers is a
statement within probability theory; its only interpretation is as a
precise mathematical statement about the probability measure on the
uncountable set of outcome sequences.

In the quantum context, we have already seen in Sec.~\ref{sec:Gutmann}
that the strong law of large numbers, applied to QPP, implies that
infinite-repetition states are eigenstates of the frequency operator,
with eigenvalues given by the absolute square of the inner product.  In
particular, when we invoke the strong law of large numbers to evaluate
the integral on the right side of Eq.~(\ref{eq:Gone}), we are
explicitly using a probability-1 statement on the uncountable sample
space of outcome sequences to define the frequency operator.  Thus,
within standard quantum mechanics, starting from QPP, a repeated
measurement on an infinite repetition state yields the frequency
eigenvalue with probability 1, not with certainty.  The point is that
PDO is not a part of standard quantum mechanics for observables on a
nonseparable Hilbert space; probability-1 predictions for measurements
of such observables do not mean that the eigenvalue occurs with certainty,
but rather can only be
interpreted as statements within probability theory.  In quantum
theory, just as for classical probabilities, we cannot interpret
probability-1 statements about infinite frequencies without reference
to an underlying notion of probabilities, and thus these statements
cannot be used to define probabilities.

An alternative to this point of view would be to assert PDO for the
frequency operator even though it is not a consequence of the
probability rule one is trying to derive.  Doing so, however, would
replace QPP not with an underlying weaker postulate, but with a strictly
stronger postulate, which would make quantum measurements different from
classical random processes, such as coin tossing, in an
incomprehensible and ultimately untestable way.

\subsection{Tail properties}
\label{sec:tailprop}

There is no problem with probability theorists' deriving purely
mathematical properties of infinite sequences of measurement outcomes
and of the infinite-copy frequency operator, for these are well defined
mathematical objects within probability theory.  The problem arises
when one tries to derive properties of finite objects from the purely
mathematical properties of infinite sequences, because there is no way
to give an operational definition of a measurement of an infinite
sequence, no way to interpret the infinite objects outside the
mathematical formalism in which they reside.  Thus, to understand
finite objects, it should not be necessary to refer to the properties
of infinite objects such as $F^\infty$.

In this subsection we pursue this line of reasoning in the quantum
setting by granting (i)~that $\ket\psi^{\otimes\infty}$ is an
eigenvector of $F^\infty$ with eigenvalue $|\langle\psi|B,0\rangle|^2$
and (ii)~the PDO conclusion that a measurement of $F^\infty$ on
$\ket\psi^{\otimes\infty}$ gives $|\langle\psi|B,0\rangle|^2$ with
certainty.  Even granting all this, we argue that one cannot reach any
conclusions about finite-copy probabilities.  The reason is that the
frequency~(\ref{eq:freqdef}) of an outcome sequence is a tail property,
which means that the limiting frequency is independent of any finite
number of initial outcomes.  It follows that any initial finite
sequence is independent of the limiting frequency.  In other words, the
fact that the limiting frequency is equal to
$|\langle\psi|B,0\rangle|^2$ is of no consequence whatsoever for the
probability of an initial finite sequence of measurement outcomes.

It is useful to emphasize precisely how this argument appears in the
quantum setting developed in Sec.~\ref{sec:infinite}.  As noted there,
{\it every\/} component contains the entire Hilbert space $\HN$ for the
first $N$ copies for any finite value of $N$; i.e.,
$\Hcomp{\psi}=\HN\otimes\Hcomp{\psi'}$, where $\seq{\psi'}$ denotes the
component's defining sequence $\seq{\psi}$ with the first $N$ vectors
omitted.  All components are identical for any finite number of copies.
Every component can thus accommodate any state for a finite number of
copies and the analysis of any measurement on those copies.  The
difference between components is wholly due to the tails of the
equivalence classes defining the components and is thus entirely irrelevant
to finite-copy considerations.

Gutmann \cite{Gutmann1995} notes that by the classical Kolmogorov
zero-one law \cite{Feller1971}, the integral over outcome sequences of
the indicator function for any tail property is equal to 0 or 1, in the
same way that the integral~(\ref{eq:quantumslln}) of the frequency
indicator function is equal to 0 or 1.  This means that we can use the
analogue of Eq.~(\ref{eq:quantumslln}) to define a quantum operator for
any tail property and that the operator has the property that
infinite-copy Hilbert-space components are eigensubspaces of the
operator.  In particular, Gutmann reports (see also
Ref.~\cite{Preskill1998c}) that Coleman and Lesniewski (unpublished)
have defined a randomness observable, $R^\infty$, such that eigenvalue
1 means that the outcome sequence has the ``Kolmogorov-Martin-L\"of
randomness property.''  In accordance with the general properties of
tail operators, Coleman and Lesniewski showed that an
infinite-repetition state is an eigenstate of $R^\infty$ with
eigenvalue 1, i.e., $R^\infty \ket\psi^{\otimes\infty} =
\ket\psi^{\otimes\infty}$, provided $|\psi\rangle$ is not one of the
eigenstates of the measured observable $B$.  From our discussion above,
it follows that this eigenvalue equation---or, indeed, the analogous
eigenvalue equation for the observable associated with any tail
property---is irrelevant to the probabilities for outcomes of any finite
number of measurements.

\section{Conclusion}   \label{sec:conclusion}

Probabilities play an important role in nearly every human endeavor.
They are central to all the sciences and especially to quantum physics,
where they appear in the very foundations of the theory.  We have our
own favorite way to interpret probabilities, the Bayesian
interpretation \cite{Bernardo1994}, which posits that probabilities
represent an agent's subjective beliefs about a set of alternatives.
This philosophical inclination has led us to propose that even quantum
probabilities are Bayesian probabilities~\cite{Caves2002a,Caves2002b}.
In the Bayesian approach, it is clear that single-trial probabilities
are a primary concept and that one {\it derives\/} properties of
frequencies in repeated trials from single-trial probabilities.

One does not have to adopt the Bayesian view---though we recommend
it---to realize that the program for defining probabilities in terms of
frequencies is bankrupt.  This paper charges an additional debt to this
bankruptcy, by showing that the inner-product structure of quantum
mechanics does not provide any additional leverage in the attempt to
derive probabilities from frequencies.  As we have seen, the quantum
frequentist program turns out to be an elaborate apparatus for
rephrasing Gleason's theorem in terms of long-run frequencies instead
of directly in terms of the quantities of interest, single-copy
probabilities.  Even this elaborate apparatus must be supported by
further unjustified assumptions that connect the desired single-copy
probabilities to infinite-copy frequencies.  Jettisoning the entire
infinite-copy apparatus allows us to deal directly with single-trial
probabilities, thus avoiding both the technical mathematics needed to
understand the nonseparable infinite-copy Hilbert space and the need
for assumptions to relate probabilities to long-run frequencies.

The final lesson of our story is clear, and it is the same for
classical and quantum probabilities: Probabilities, not frequencies,
are the primary concept; inferences always run not from frequencies to
probabilities, but from probabilities to statistical properties of
frequencies.

\acknowledgments

We profited from discussions with J. B. Hartle. This work was supported in
part by the U.S. Office of Naval Research Grant No.~N00014-03-1-0426.

\end{document}